\documentclass[]{spie}  
\usepackage[]{graphicx}

\title{Characterizing Near-Infrared Sky Brightness in the Canadian High Arctic} 

\author{Suresh Sivanandam\supit{a}, James R. Graham\supit{a}, Roberto Abraham\supit{b}, Anthony Tekatch\supit{c}, Eric Steinbring\supit{d},   Wayne Ngan\supit{b}, Doug L. Welch\supit{e,c}, Nicholas M. Law\supit{a}
\skiplinehalf
\supit{a}Dunlap Institute, University of Toronto, 50 St. George St, Toronto, ON, Canada; \\
\supit{b}Department of Astronomy, University of Toronto, 50 St. George St, Toronto, ON, Canada; \\
\supit{c}Unihedron Inc., 4 Lawrence Ave., Grimsby, ON, Canada; \\
\supit{d}National Research Council Canada, 5071 West Saanich Road, Victoria, BC, Canada; \\
\supit{e}Department of Physics and Astronomy, McMaster University, Hamilton, ON, Canada;
}

\authorinfo{Further author information: (Send correspondence to S.S.)\\S.S.: E-mail: sivanandam@di.utoronto.ca, Telephone: +1 (416) 978-2183}

\begin{document} 
  \maketitle 

\begin{abstract}

We present the first measurements of the near-infrared (NIR), specifically the $J$-band, sky background in the Canadian High Arctic. There has been considerable recent interest in the development of an astronomical observatory in Ellesmere Island; initial site testing has shown promise for a world-class site. Encouragement for our study came from sky background measurements on the high Antarctic glacial plateau in winter that showed markedly lower NIR emission when compared to good mid-latitude astronomical sites due to reduced emission from the Meinel bands, i.e. hydroxyl radical (OH) airglow lines. This is possibly a Polar effect and may also be present in the High Arctic. To test this hypothesis, we carried out an experiment which measured the the $J$-band sky brightness in the High Arctic during winter. We constructed a zenith-pointing, $J$-band photometer, and installed it at the Polar Environment Atmospheric Research Laboratory (PEARL) near Eureka, Nunavut (latitude: 80$^\circ$ N).  We present the design of our ruggedized photometer and our results from our short PEARL observing campaign in February 2012. Taken over a period of four days, our measurements indicate that the $J$-band sky brightness varies between $15.5-15.9$ mag arcsec$^2,$ with a measurement uncertainty of 0.15 mag. The uncertainty is entirely dominated by systematic errors present in our radiometric calibration. On our best night, we measured a fairly consistent sky brightness of $15.8\pm0.15$ mag arcsec$^2.$ This is not corrected for atmospheric extinction, which is typically $<0.1$ mag in the $J$-band on a good night. The measured sky brightness is comparable to an excellent mid-latitude site, but is not as dark as claimed by the Antarctic measurements. We discuss possible explanations of why we do not see as dark skies as in the Antarctic. Future winter-long sky brightness measurements are anticipated to obtain the necessary statistics to make a proper comparison with the Antarctic measurements. 

\end{abstract}



\section{INTRODUCTION}
\label{sec:intro}  
\begin{figure}
   \begin{center}
   \begin{tabular}{c}
   \includegraphics[height=8cm]{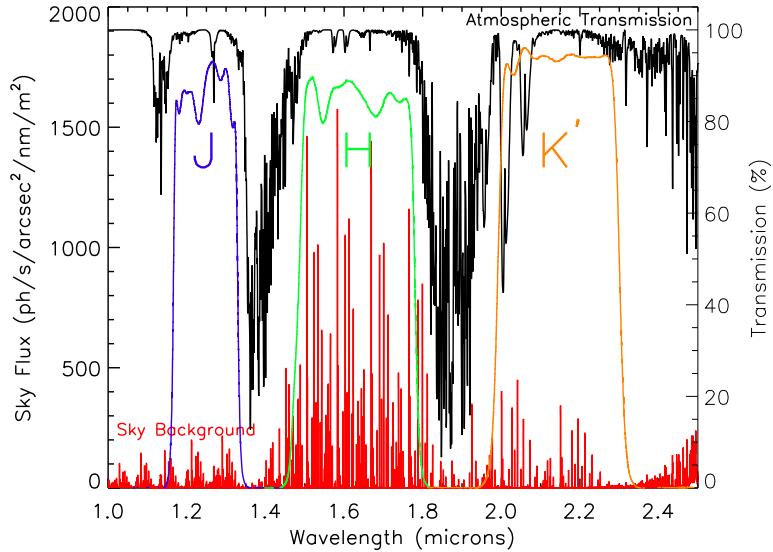}
   \end{tabular}
   \end{center}
   \caption[nearirsky] 
   { \label{fig:nearirsky}  The NIR atmospheric transmission for an excellent mid-latitude site, shown here for Mauna Kea (black curve) and over plotted with a corresponding high-resolution spectrum of the sky background (red). The purple, green, and orange curves are the Mauna Kea Observatories (MKO) $J$, $H$, and $K^\prime$ filter bandpasses. Most of the sky flux in this wavelength regime is completely dominated by OH airglow lines. Only at the very end of the NIR band does the thermal emission from the sky start to dominate. (Colour version available online) 
}
   \end{figure}

The sky background plays a significant role in astronomy as it affects all forms of astronomical observations across the entire electromagnetic spectrum. The sky background in space and on the ground from the far ultraviolet to the far infrared wavelengths is well studied, and it is now known that a multitude of emission processes dominate different wavelength regimes\cite{leinert98}. The characterization of the sky background is crucial in astronomy because the sky brightness often determines the length of the time required to reach a given signal-to-noise for an astronomical observation. This is especially the case in the near-infrared (NIR) where the sky background is significantly brighter than in the visible. The NIR bands are mostly dominated by non-thermal emission, namely very narrow airglow lines that have been identified to be hydroxyl radical (OH) lines, Meinel bands \cite{meinel50,meinel50b}, and to a lesser extent O$_2$ ($^1 \Delta _g$ band) emission at 1.27 $\mu$m\cite{maihara93}. The airglow lines are generated by the vibrational de-excitation of excited OH molecules created in the mesosphere at an altitude of $80-105$km\cite{ramsay92}. The main reaction that produces this molecule is the following:
\begin{equation}
\textrm{H}+\textrm{O}_3\rightarrow\textrm{OH}^*+\textrm{O}_2,
\end{equation}
where OH is formed in a vibrationally excited ($\nu \le 9$) state\cite{bates50}. The excited OH radical cascades radiatively to lower energy states and produces the characteristic airglow spectrum observed in the NIR (see Figure \ref{fig:nearirsky}). This emission is highly variable in short (minutes) and longer (hours to years) timescales. The short time scale variations of tens of minutes have been identified to be caused by the passage of gravity (buoyancy) waves high in the atmosphere that modify the densities of reactants that produce the OH molecule\cite{ramsay92,swenson94}. Longer time scale diurnal variations along with correlations with the solar cycle have been observed. There have been studies that show that the OH airglow is brighter during periods of intense solar activity\cite{pertsev08}. 

Significant work has been done to find ways to suppress this OH background and improve NIR sensitivity. These techniques require complex hardware to accomplish this task\cite{herbst94,hawthorn11}. The goal of these methods is to reach the IR continuum in between the OH lines, which is significantly fainter and dominated by the zodiacal background though there may be other emission mechanisms at work\cite{maihara93}. Another alternative to reduce the OH background component is to find a site that has intrinsically lower OH background. A winter campaign carried out at the South Pole to measure the IR sky background from the near to mid-IR revealed that the sky background was significantly fainter than a mid-latitude site\cite{phillips99}. In the middle of winter at the South Pole, there should be continuous darkness resulting in little to no diurnal variations in the sky brightness. It is not particularly surprising that the thermal sky emission was measured to be an order of magnitude or more fainter than a comparable mid-latitude site because of the frigid temperatures ($\sim -60$C) and very low precipitable water vapour at the South Pole. But the $J$ and $H$-band sky background values were also fainter by factors of 2 to 3 ($\sim1$ mag fainter). This is particularly intriguing because it suggests the chemical process that produces the Meinel bands in the South Pole region is somehow less effective. 
\par
Recent investigation of a potential Canadian astronomical site near Eureka, Nunavut at 80$^\circ$ N\cite{steinbring12}, and others on Ellesmere Island at higher latitude\cite{steinbring10}, prompted us to test the hypothesis that Arctic skies may also experience suppressed OH airglow in winter. There are a number of virtues of Eureka. Being located very close to the North magnetic pole\footnote[1]{The magnetic pole is not stationary and moves a few 10s of km per year.}, Eureka is within the auroral hole. It also remains accessible throughout the winter through chartered flights. Finally, the environment is not as harsh as the South Pole as average winter temperatures hover around $-40$C, the minimum rated temperature for most electronic equipment. Our primary objective was to carry out a long term monitoring of the NIR sky brightness of the site. We chose the Polar Environment Atmospheric Research Laboratory (PEARL), at an elevation of 610-meters, located approximately 15 km from the Eureka station, as the location of our site testing campaign due to both the availability of electrical power and internet access. We focussed primarily on the $J$-band as that was where the greatest improvement in the sky background was observed in the $1-2 \mu$m wavelength range in Antarctica. For this purpose, we constructed a ruggedized $J$-band photometer that monitored the sky brightness continuously. There was a short site testing campaign that took place in February 2012 when the photometer was set up for 11 days. During this time, we obtained four nights of good sky brightness data. In this paper, we present the results of the first sky brightness measurements in the High Arctic, along with the design, construction, and calibration of the photometer. We conclude with the issues we experienced and our future plans for NIR sky brightness characterization. 

   \begin{figure}
   \begin{center}
   \begin{tabular}{c}
   \includegraphics[height=6cm]{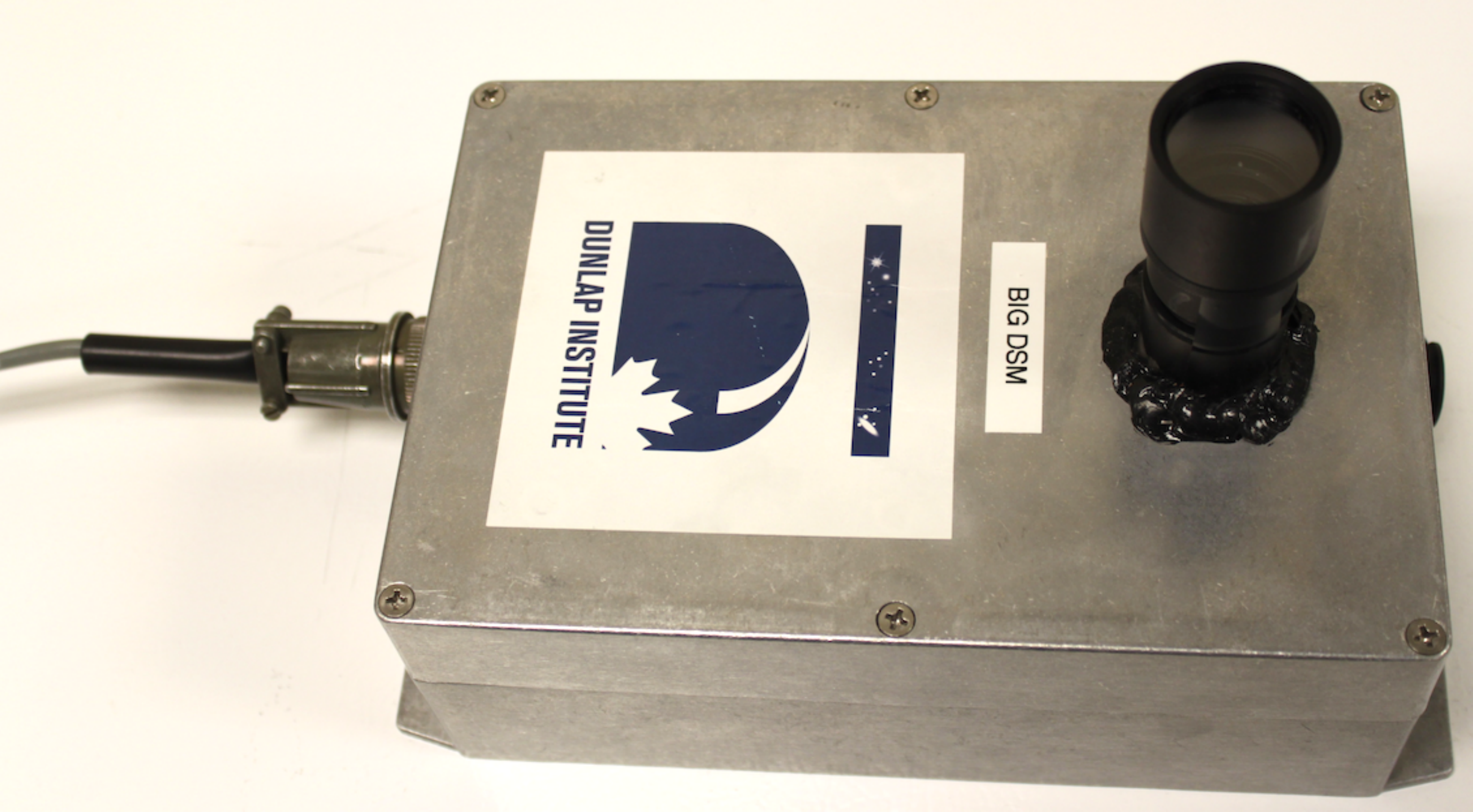}
   \end{tabular}
   \end{center}
   \caption[photometer] 
   { \label{fig:photometer} A photograph of the $J$-band photometer. The black lens tube contains all of the optical components of the photometer. The electronics and the photodiode are housed in a sealed, protective box. Both power and communication are via the cable at left. There is a MIL-SPEC connector for the cable on one end of the photometer that provides a good environmental seal.
}
   \end{figure} 

\section{Instrument Design} 
\label{sec:design}
The goal of the design was to construct a simple photometer that can withstand temperatures reaching $\sim -40$C while being able to measure the zenith $J$-band sky brightness at a high signal-to-noise ratio (SNR). Eureka is a somewhat more forgiving environment than the South Pole because most common electronic components are rated to operate at temperatures down to $-40$C. Nevertheless, one needs to be mindful of using movable components. We already have significant experience constructing precise current measuring circuits that are central to commercial single pixel, visible-light sky brightness monitors constructed by our co-authors. These devices have been shown to meet the necessary operating temperature requirement. We used this experience to construct an infrared sky brightness photometer sensitive enough to accurately monitor the $J$-band sky with minimal moving parts. In the interest of cost, reliability, and ease of construction, we chose a zenith-pointing, single pixel photometer design with no moving parts. 
\par
There were a number of design considerations that came into play in the construction of the photometer. The $J$-band passband was chosen for a number of reasons: (1) The $J$-band sky background is entirely dominated by OH airglow; (2) The greatest reduction in OH airglow-dominated sky brightness in Antarctica was seen in $J$-band\cite{phillips99}; and (3) The typical InGaAs diode sensitivity cuts off at $1.7\mu$m, which means part of $H$ band will not be accessible. Extended wavelength diodes can be purchased, but they are significantly more expensive, and have lower spectral detectivity, D$^*,$ values. The next design consideration is the accuracy to which we could measure the photocurrent generated by sky photons. From previous experience we knew that our current sensing circuits could measure currents with reasonable accuracy of about 0.25 pico amp (pA). We set a design requirement of sky generated photocurrent of $\sim5$ pA. This requirement drives the field-of-view (FOV) and aperture size of the photometer. Because we intend to measure the flux from an extended source, we need to maximize the photometer's etendue, i.e. $A\Omega$ where $A$ is the collecting area, and $\Omega$ is the solid angle of our photometer. 
\par
\begin{figure}
   \begin{center}
   \begin{tabular}{c}
   \includegraphics[height=5cm]{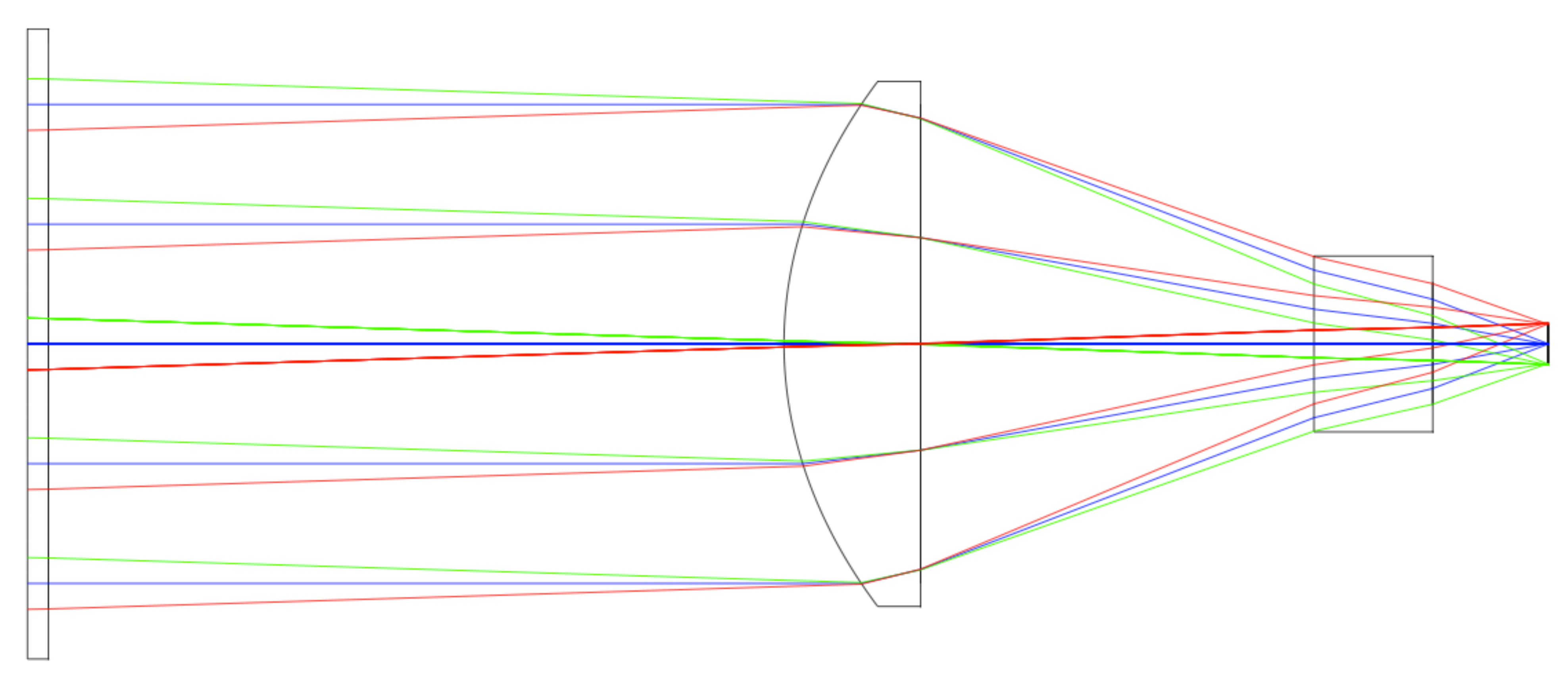}
   \end{tabular}
   \end{center}
   \caption[layout] 
   { \label{fig:layout} 
   The optical layout of the $J$-band photometer. Light enters from the left and the different coloured rays show the rays from different field angles. The red and green rays define the edge of the photometer's FOV. The flat sapphire window is shown to the left, followed by the aspheric lens, the $J$-band filter, and the focal plane. The stop of the system is located at the rear of the aspheric lens, which is defined by the inner diameter of the lens tube. The lens is mounted on a focusing stage, which is used to set the proper focus during assembly. The distance from the sapphire window to the lens is 35 mm. (Colour version available online) 
   }
   \end{figure}

For our first design we considered a simple radiometer, which consisted of a baffle tube with a photodiode at one end. In this case, $A$ is set by the size of the photodiode, which was 2mm in diameter, and $\Omega$ is defined by the baffle tube, which was approximately 15 degrees. We constructed a prototype of this design and found that one needed a larger FOV to reach a sufficient signal-to-noise ratio. A larger FOV would mean greater sensitivity to variations in sky brightness and atmospheric extinction across the sky. The bigger issue was that the angular response of the radiometer varied strongly with of field position and required fairly accurate lab characterization to determine the solid angle of the device. Our final revised design made use of a lens, which improved the sensitivity over the original baffle-tube prototype by a factor of 10. The lens radiometer provides a well-defined field-of-view with a uniform angular response across its FOV. In order to maximize the etendue, the lens needs to be fairly fast, i.e. low f/\#. This drove us to an aspheric lens design that used an InGaAs photodiode with a relatively large active area. For our lens, we used an off-the-shelf Edmund Optics precision asphere lens (part \#49115), which is 1-inch in diameter and has a focal length of $\sim30$mm. The lens is antireflection-coated to be less than 1.5\% reflective in the $0.6-1.05$ $\mu$m range. The diode was also an off-the-shelf part from Hamamatsu Corp (part \#G8605-22). The diode has a two stage thermoelectric cooler and an active area 2 mm in diameter. The $J$-band filter was obtained from Optec, which makes these filters for IR photometers. The $J$-band filter closely resembles the Mauna Kea Obsevatories (MKO) $J$-band filter\cite{tokunaga02}. A 30-mm uncoated Sapphire entrance window was used to protect the optical components from the elements. The optical layout of the system is shown in Figure \ref{fig:layout}. The optical parameters of the photometer are given in Table \ref{tab:parameters}. 

 \begin{table}[h]
\caption{System parameters for $J$-band photometer} 
\label{tab:parameters}
\begin{center} 
\small      
\begin{tabular}{|l|l|} 
\hline
\rule[-1ex]{0pt}{3.5ex}  {\bf Aperture Diameter} & 20.8 mm \\
\hline
\rule[-1ex]{0pt}{3.5ex}  {\bf f/\#} & 1.5\\
\hline
\rule[-1ex]{0pt}{3.5ex}  {\bf Field-of-view } &  3.5 degree diameter\\
\hline
\rule[-1ex]{0pt}{3.5ex}  {\bf Sky Area (Solid Angle)} & $1.32\times10^8$ arcsec$^2$   \\
\hline
\rule[-1ex]{0pt}{3.5ex}  {\bf Filter Center Wavelength} & 1.260 $\mu$m  \\
\hline
\rule[-1ex]{0pt}{3.5ex}  {\bf Filter Bandwidth (Full-Width-Half-Max)} & 0.1292 $\mu$m   \\
\hline
\rule[-1ex]{0pt}{3.5ex}  {\bf Filter Throughput} & $\sim 0.8$ \\
\hline
\rule[-1ex]{0pt}{3.5ex}  {\bf End-to-end Throughput (predicted/measured)} & $\sim 0.67/\sim 0.7$ \\
\hline
\rule[-1ex]{0pt}{3.5ex}  {\bf Average Responsivity} & $\sim0.84$ A/W   \\
\hline
\rule[-1ex]{0pt}{3.5ex}  {\bf Predicted Sky Photocurrent} & $\sim 4$ pA @ 16.0 mag arcsec$^{-2}$ \\
\hline
\end{tabular}
\end{center}
\end{table}

\par
\begin{figure}
   \begin{center}
   \begin{tabular}{c}
   \includegraphics[height=8cm]{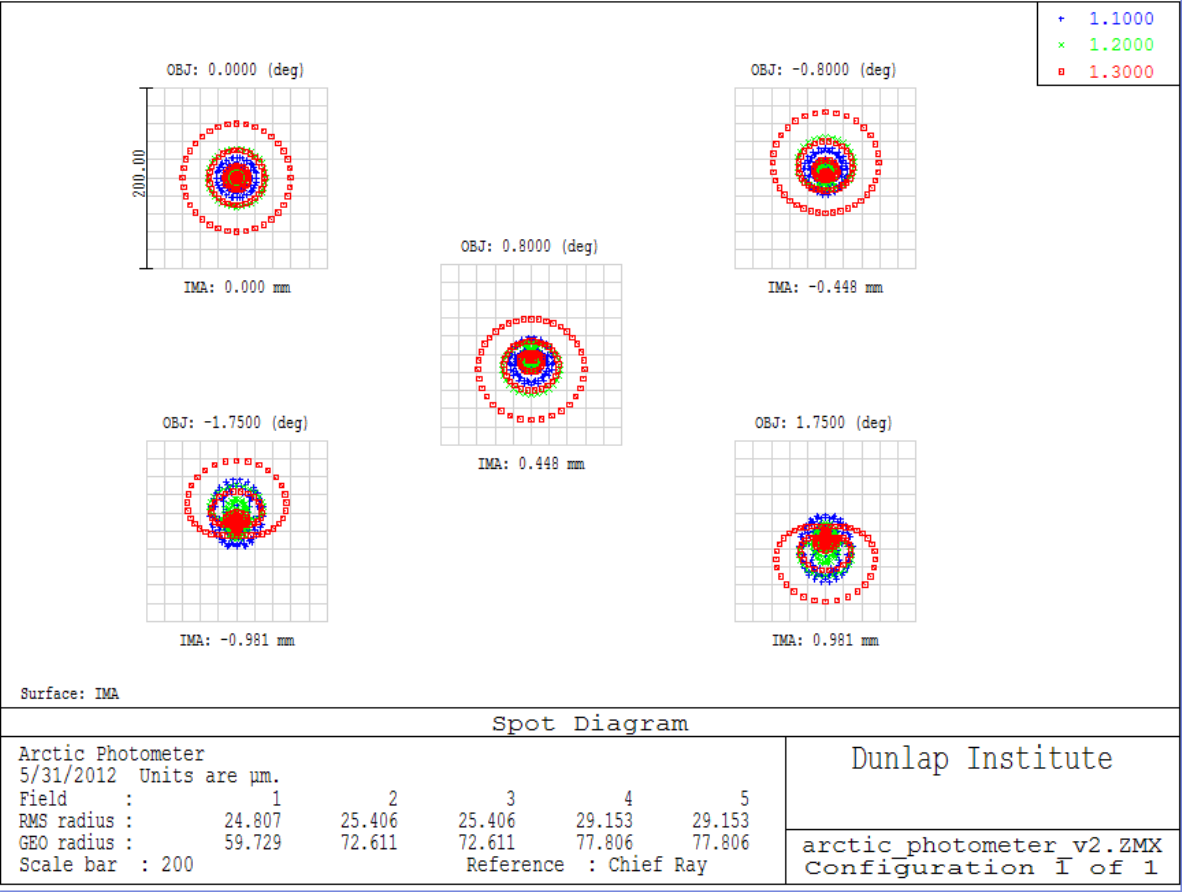}
   \end{tabular}
   \end{center}
   \caption[spotdiagram] 
   { \label{fig:spotdiagram} The spot diagram for the aspheric photometer design over multiple fields. The blue, green, and red colours represent the spots generated by rays at wavelengths of 1.1, 1.2, and 1.3 $\mu$m, respectively. These wavelengths span the $J$-band bandpass. The spot diagram shows that the performance is fairly uniform across the full field at $J$-band wavelengths. RMS spot radii are typically in the $25-30$ $\mu$m range. (Colour version available online) }
   \end{figure}

\begin{figure}
   \begin{center}
   \begin{tabular}{c}
   \includegraphics[height=6cm]{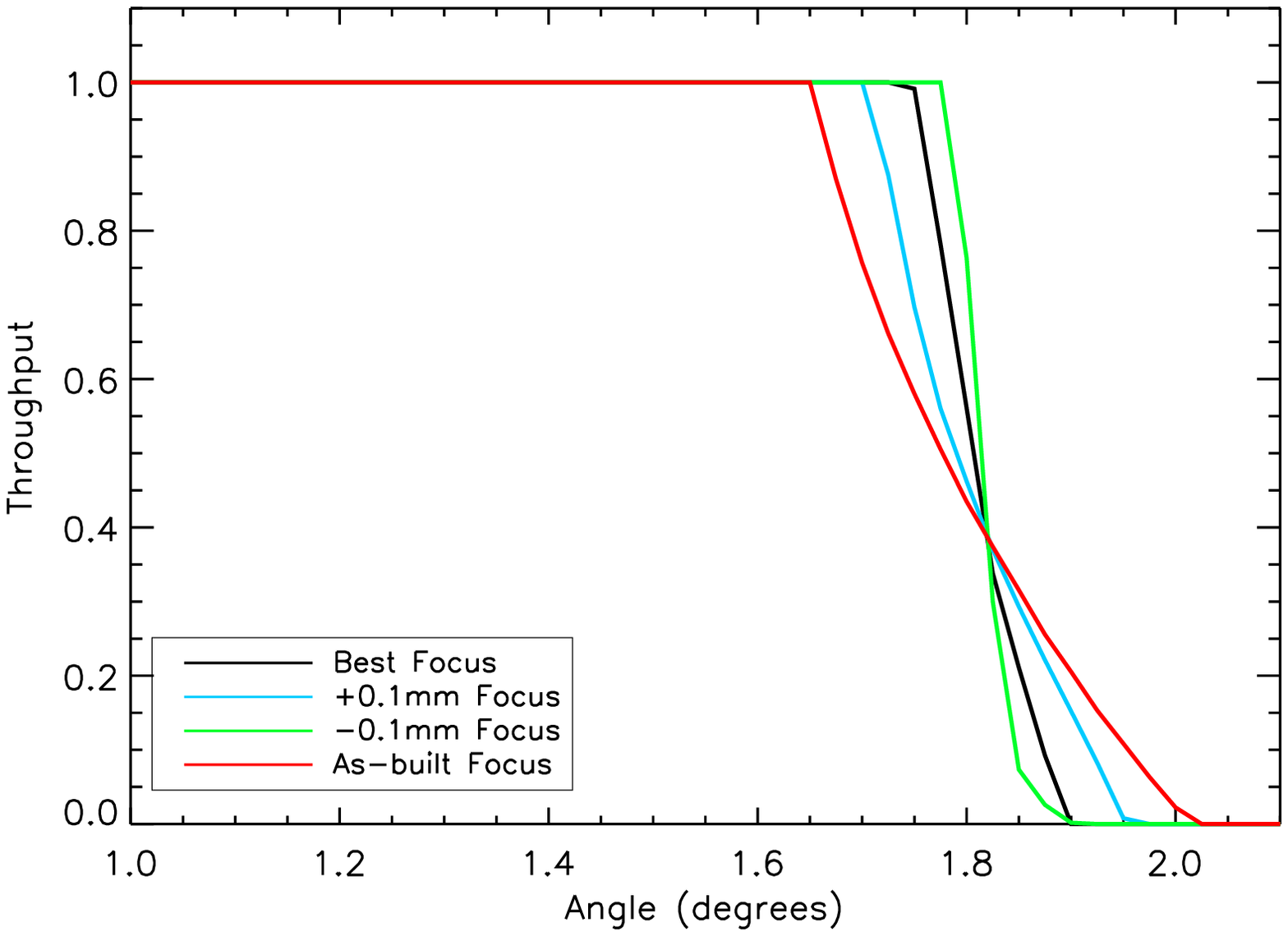} \includegraphics[height=6cm]{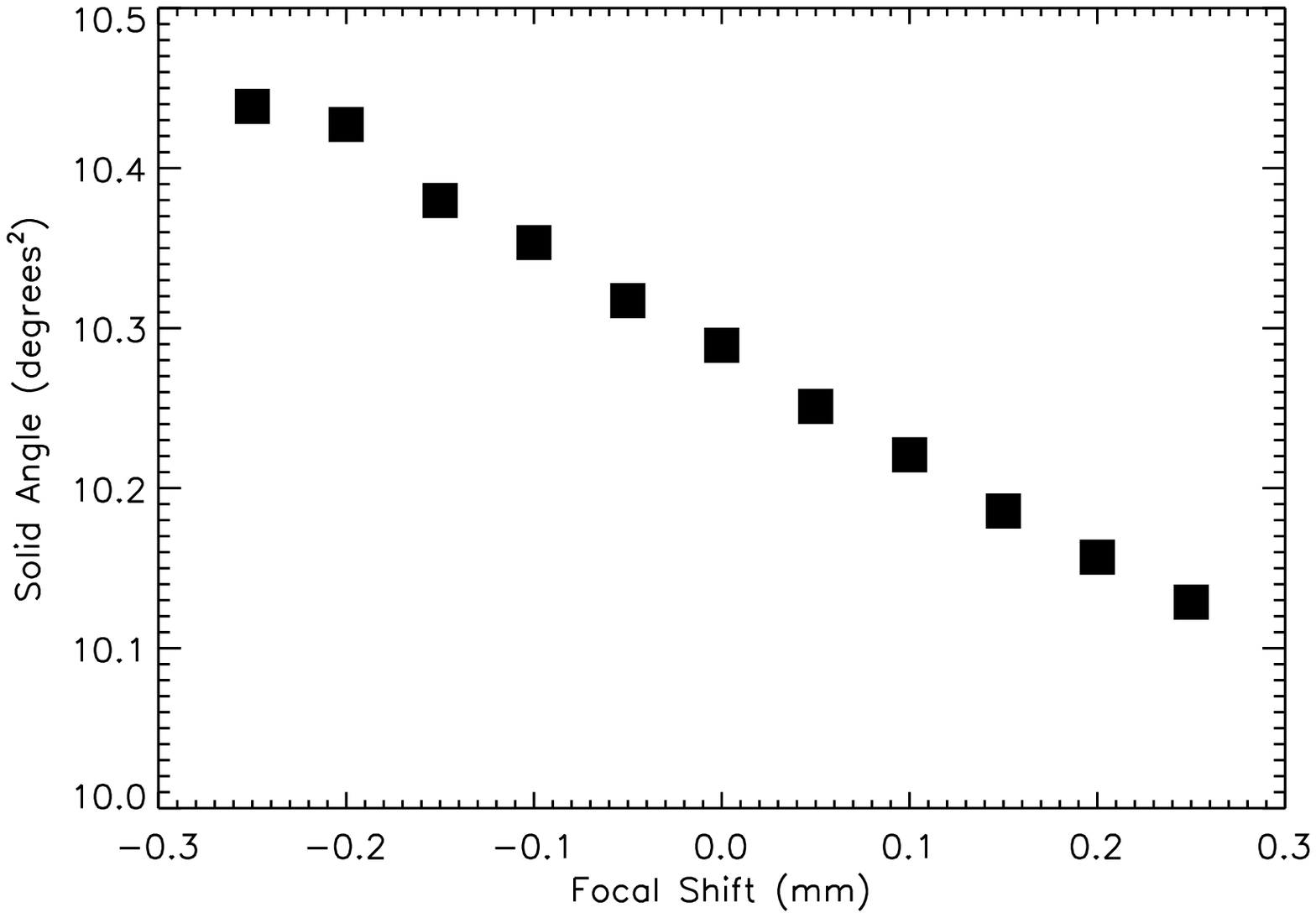}
   \end{tabular}
   \end{center}
   \caption[fov] 
   { \label{fig:fov} {\it Left:} Simulated angular response function for the photometer for different focus settings of the photometer. The throughput at angles smaller than 1 degree is 1.0. This region has not been plotted to accentuate the changes in the response function at larger angles. The nominal case is shown by the black curve where the lens is properly set to infinity focus. The blue and green curves show the effect of a slight focal shift on the angular response function. The red curve shows the as-built simulated angular response function. {\it Right:} Solid angle of photometer as a function of focal shift of lens. 0 mm is defined to be nominal infinity focus case. The as-built focal position of the photometer is expected to be +0.2 mm from nominal. In all cases, the solid angle of the photometer does not change more than a few percent from the infinity focus case. (Colour version available online) }
   \end{figure}

 
The goal of the optical design was to obtain relatively uniform angular response, while maximizing throughput, over the full FOV of the photometer. This requires fairly small root-mean-squared (RMS) spot sizes that are much smaller than the size of the detector. This ensures all of the focussed light falls on the detector over the majority of the photometer's FOV. Conventional fast spherical lenses exhibit significant spherical aberration and are unable to satisfy this requirement. The aspheric design works remarkably well, and its spot diagram over the full field is shown in Figure \ref{fig:spotdiagram}. Typical RMS spot sizes are within $25-30$ $\mu$m across the entire field over the wavelengths of interest. We simulated the angular response function for and solid angle subtended by the photometer using ZEMAX. The results are shown in Figure \ref{fig:fov}. For the ideal case when the photometer is properly focussed at infinity, the angular response function (ARF) has a fairly sharp transition at 1.75 degrees as expected for our design. This is because the RMS spot sizes are much smaller than the detector size. We also show the effects of an improperly set focus and how that affects the ARF in the left panel of Figure \ref{fig:fov}. Small changes of $\pm 0.1$mm, which is much coarser than the our ability to set the lens separation, has a minor effect on the ARF. The solid angle subtended by the photometer for the slightly defocussed case does not change appreciably from the perfectly focussed case, see Figure \ref{fig:fov} right panel. However, we were unable to focus the photometer at infinity. Instead, the focussing was carried out by trying to maximize the observed signal of a fairly collimated ($\pm 5$ degrees), IR ($\lambda = 950$ nm) LED light source located 3-meters away. This means that the focus is not properly set to infinity, and the effect on the ARF is shown in Figure \ref{fig:fov} as the red curve. However, the solid angle of this as-built setup only differs by 2\% from the nominal infinity focus case, which means the focal shift does not significantly impact our knowledge of the photometer's solid angle (given in Table \ref{tab:parameters}). From this point on, we will use this solid angle to be the nominal solid angle of the photometer.
\par
We estimated the expected photocurrent for our $J$-band sky brightness measurements. Assuming the optimistic case of a sky brightness of 16.0 mag arcsec$^{-2},$ which was on the upper end of the typical sky brightness values measured in the Antarctic \cite{phillips99}, we estimate a photocurrent of 4 pA. This assumes a total photometer throughput of $\sim 0.67$ where we have a sapphire window transmission of 0.86, a lens transmission of 0.97, and filter transmission of 0.8. The diode responsivity is assumed to be 0.84 A/W, which was calculated by taking a weighted average value of a typical Hamamatsu InGaAs photodiode responsivity curve across the $J$-band bandpass. For the most optimistic case of a sky brightness of 16.8 mag arcsec$^{-2},$ the lower end of the Antarctic measurements, we would measure a photocurrent of 2 pA. Our photometer should detect this signal at the 5$\sigma$ level.

 \par
The final component of the photometer is the photocurrent measurement device. The photocurrent is measured by a custom-built electronic circuit that accomplishes the following tasks: (1) Measures and digitizes the photocurrent; (2) Allows one to change the gain and integration time for these measurements; and (3) Communicates with a computer over a serial line. To communicate with the photometer, we use a netbook computer that executes a PERL script to setup the photometer's operating parameters and retrieve data. A protocol that uses a suite of text commands is used to communicate with the photometer. The netbook then creates a text file of the measurements that can be retrieved at later time via an Internet connection.
\par
The heart of the custom circuit consists of a current input 20-bit analog-to-digital converter (ADC) - Texas Instruments DDC112 - and a PIC microprocessor, which provides a bidirectional link to a computer over a serial line and sets up the DDC112 to carry out the current measurements. The photodiode is configured in zero bias mode, which means there is no dark current. We do not require high speed measurements, which makes the lack of biasing a worthwhile compromise because at zero bias the diode response to light is very linear over a wide intensity range. The DDC112 is able to make unipolar current measurements to a high degree of accuracy over a wide temperature range. Schematically, the DDC 112 is an integrator with a known capacitance (5\% tolerance) and its output is fed into a built-in ADC.   There are several integration capacitors that can be chosen depending on the amount of charge that needs to be measured for a given integration time, which essentially changes the gain of the measurement. After one integration time, the digitized photocurrent value is retrieved by the microprocessor and can be read by the control computer. In our case, due to the exceedingly small photocurrent to be measured, we use the smallest integrating capacitor (12.5 pF) and the longest possible integration time of one second. The full well charge level at this setting is 50pC, which means we can measure a current up to 50 pA at 1s integration time without saturating the integrator. At this setting, we estimate a 1$\sigma$ current measurement of 0.25 pA. This error estimate is obtained by characterizing the stability of measurements and shifts in the bias values as a function of temperature. This means that the predicted photocurrent of $\sim 4$ pA is detected at a high level ($> 10\sigma$) of significance. 
\par
Another function that we implemented in the circuit is the temperature control of the photodiode. The photodiode in the photometer has a thermoelectric cooler (TEC) that can be used to cool the diode 40C below ambient. This feature improves the stability of the photodiode and reduces the leakage current, see Section \ref{sec:issues} for further discussion. The circuit also consists of a temperature sensing circuit that determines the diode temperature by measuring the resistance of its internal thermistor. The TEC driver is a constant current source that is turned on and off by a thermostat control implemented in the photometer firmware. This implementation is able to maintain the temperature of the diode to within 1C.
   
\section{Calibration and Testing}
\label{sec:calibration}
In order to know the conversion between counts measured by the photometer and the true sky brightness, one needs to know the flux calibration of and the solid angle subtended by the photometer. The ZEMAX simulations show that the solid angle of the photometer is insensitive to slight errors in the focus adjustment. We did carry out crude measurements of the angular response that were not at the level necessary to confirm the solid angle of the photometer to high accuracy, but we found our measurements to be consistent with simulations. It is possible that there might be additional sensitivity at larger angles due to scattering inside the lens barrel. This will only increase the solid angle of the photometer, and therefore reduce the total sky brightness reported by the photometer. Given that our goal was to determine if the High Arctic was as dark as the South Pole, our measurements would then set an upper limit on the measured sky brightness. We estimate a 10\% uncertainty in our solid angle determination. Our calibration procedure focused entirely on the flux calibration of the photometer. 
\par
The flux calibration of the photometer was done in a lab environment using a calibrated light source. It is not possible to carry out a flux calibration by observing an astrophysical source since even the brightest stars are too faint, and the Sun is too bright. The solution then was to use a calibrated lab source to carry out the flux calibration. The lab characterization setup used a 1310 nm pigtailed laser with a collimator at its fibre end. The 1310 nm laser is ideal because it is within the $J$-band bandpass. The output of the collimator produced a Gaussian beam with a measured beam waist of 1.5 mm and a divergence angle of $\sim0.3$ degrees. The laser was set to operate in constant power mode, and an InGaAs photodiode 2mm in diameter with a calibrated response curve that was NIST traceable was used to measure the laser's power output by placing the photodiode very close to the collimator output. An ND filter was inserted at the output of the laser collimator to attenuate the light to a level that did not cause the photodiode readout to saturate at the lowest gain and shortest integration time. To determine what fraction of the light from the laser fell within the diode active area, we scanned the diode across the laser beam and recorded its readout as a function of scan distance. The best fit to the scan curve indicated that $95-97$\% of the laser light fell within the diode active area. To test the photometer, we kept the ND filter in the beam and set up the photometer such that the optical axis of the collimated laser beam and the photometer optics were co-aligned.  Coalignment was thought to have been achieved when the highest readout was obtained from the photometer while the beam was centred on its lens. Note that the width of the laser beam was only 5 mm at the photometer entrance aperture and we did not fill the full clear aperture. Nevertheless, all of the light incident on the photometer should be focussed on the photodiode, and one should be able to obtain a flux calibration of the system. This calibration obtained a throughput of $\sim 0.7,$ which assumes the responsivity of the calibration diode and the photometer diode are the same at 1.31 $\mu$m. This measurement should be to good to within a few percent. This is also roughly consistent with our predicted throughput for the system. Based on the stability of our measurements, we expect this flux calibration to be good to 10\%. The weakness of this calibration technique is that it does not test the photometer at the same flux levels and spectrum as the sky background, and the same environmental conditions as in the Arctic. Furthermore, the calibrated light source does not fill the full aperture of the photometer, which means spatial variations in transmission of the optical system and the responsivity of the diode can affect our calibration. More careful calibrations need to be done in the future to verify the accuracy of this method. The total uncertainty in the sky brightness measurement, dominated entirely by systematic errors in the photometer's flux calibration and solid angle determination, is therefore $\sim 15\%$ or 0.15 mag. 
\par
The system was tested in the dark to determine its bias value. We discovered that its bias shifted significantly as a function of temperature. This was already discovered when the first prototype was constructed, which did not have a thermoelectric cooler. The photometer's onboard temperature measurement was used to correlate the changes in the bias value. While a general trend was discovered, it was difficult to accurately predict the expected bias value for a given temperature. Further testing discovered that two main components affected the bias value of the diode: (1) A leakage current flowing in the opposite direction of the photocurrent that was highly sensitive to temperature; (2) Thermal flux detected by the photodiode from emission from the optical and mechanical components near it; and (3) The operation of the thermoelectric cooler. The first issue can be mitigated by cooling the diode, but the second two are intrinsic to the photometer design. We discuss these issues further in Section \ref{sec:issues}. Fortunately, none of these issues are of concern at the very low temperatures the photometer operates at in the High Arctic where active cooling is not necessary. However, they are of considerable concern for operation in warmer temperatures if, for example, the photometer used to test a warmer mid-latitude site. 

\section{High Arctic Sky Background Measurement}
\label{sec:irskyback}
\begin{figure}
   \begin{center}
   \begin{tabular}{c}
   \includegraphics[height=12cm]{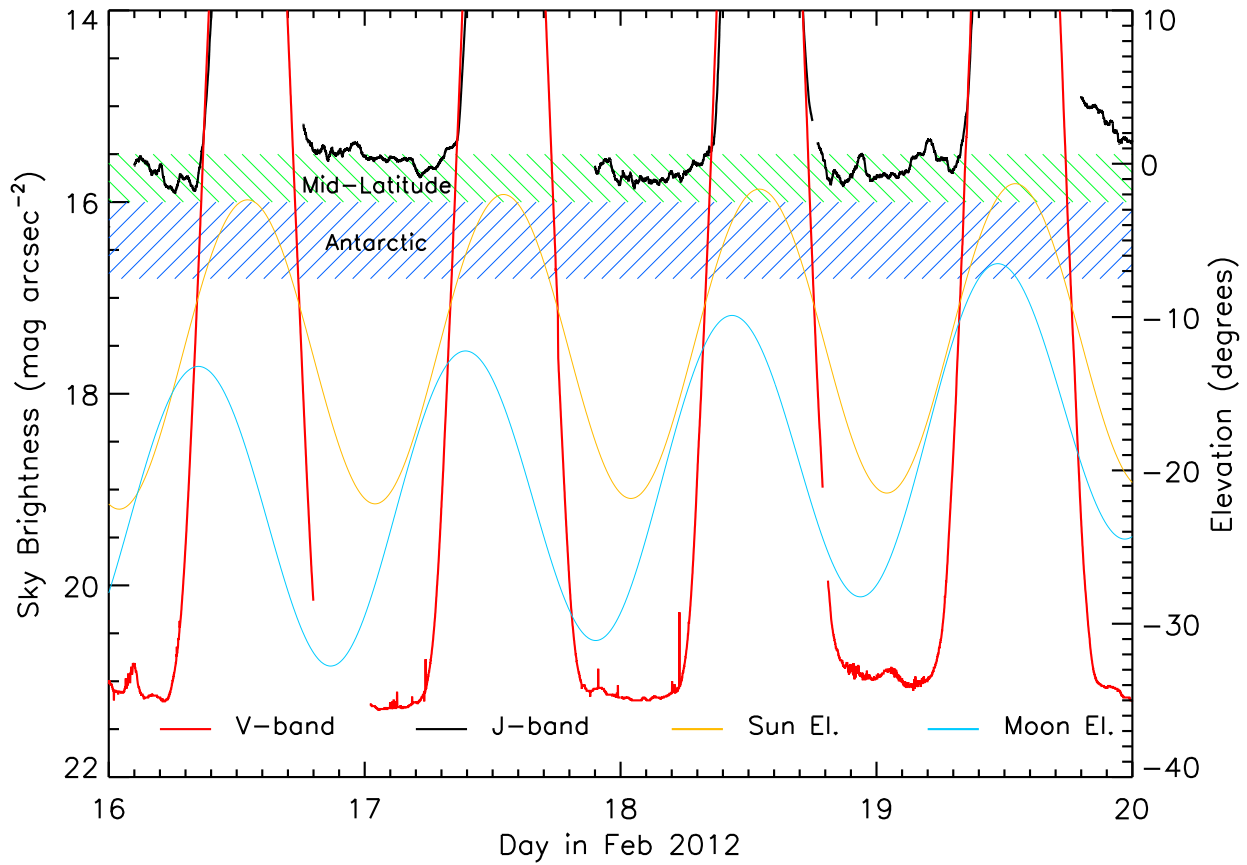}
   \end{tabular}
   \end{center}
   \caption[skyback] 
  {The measured sky background over a four day period at the PEARL station near Eureka, NU. Time is shown in Eastern Standard Time (EST). The black curve represents the $J$-band sky brightness, while the red curve represents the $V$-band sky brightness. The gaps in the data streams are regions with unreliable data, i.e. snow covered windows or spikes in brightness due to personnel moving near the photometers. The orange and light blue curves represent the Sun and Moon elevations during that time. Neither the Sun nor the Moon were above the horizon during the period, but the Sun did spend a significant fraction of the time above $-12^\circ$ twilight. The green hatched region represents typical Mauna Kea values while the blue hatched region cover the typical range of values observed at the South Pole\cite{phillips99}. (Colour version available online) \label{fig:skyback} }
   \end{figure}

We measured the $J$-band sky brightness at PEARL over a period of 11 days during which we continuously operated the photometer. During this period, we obtained four good days of data. Some days were lost due to poor weather depositing snow on the entrance window while others were clouded out. We also installed an off-the-shelf Unihedron visible light sky quality meter (Model: SQM-LU), which measured the $V$-band sky brightness at the site and served as an additional source of information. The $V$-band sky monitor had a larger field that was 20 degrees (half-power beam width). The IR measurements were taken in 10 second intervals, while the $V$-band measurements were taken in one minute intervals. We show our results in Figure \ref{fig:skyback}. During this period of four days, both the Moon and Sun remained below the horizon producing the darkest skies when the Sun was below $-18^\circ$ elevation. The $J$-band sky brightness varied from 15.5 to 15.9 mag arcsec$^{-2}$ during this time. On the best night, which stretched from the 17th to the 18th, the $J$-band sky brightness remained consistently at 15.8 mag arcsec$^{-2}.$ This is not corrected for atmospheric extinction, which is typically $< 0.1$ mag in $J$-band at a good astronomical site\cite{lombardi11}. Also, the $V$-band sky brightness values ranged within 21.0 to 21.3 mag arcsec$^{-2}$ during this period. If we account for a typical $V$-band atmospheric extinction of 0.2 mag, these values agree with previous measurements\cite{steinbring12}. Both of these measurements suggest that the site is of excellent quality in terms of sky brightness. The $J$-band sky brightness is consistent with typical values at the best mid-latitude astronomical sites, but is still brighter than the Antarctic measurements. A compilation of NIR sky-brightness values for mid-latitude astronomical sites is available elsewhere\cite{sanchez08}.

\par
There could be a number of reasons why our measurements are higher than at the South Pole; one being that the OH airglow is not suppressed in the High Arctic. More data are needed to be convinced of that though, as our statistics are not yet robust. The Antarctic measurements were carried out continuously over a course of four winter months; however, we only obtained approximately four days of good data. While the majority of the Antarctic $J$-band sky brightness measurements fall within the $16.0-16.8$ mag arcsec$^{-2}$ range, there is a long tail on the bright end that stretches out to 15.5 mag arcsec$^{-2}$\cite{phillips99}. Our current sky brightness measurements do not have enough statistical significance to suggest that we are comparable or worse than the Antarctic measurements. A longer term monitoring program is required. Another explanation may be one related to solar activity. It has been shown that the OH airglow brightness is correlated with solar activity measured in the form of 10.7-centimeter (cm) solar microwave flux\cite{pertsev08}. It was found that the OH airglow varied by $30-40$\% per 100 solar flux units (sfu) measured at 10.7cm\cite{pertsev08}. The original South Pole measurements were taken close to the solar minimum during 1995. However, our measurements were taken in 2012 as the Sun was approaching its maximum period of activity. We obtained the sfu measurements for both periods from Space Weather Canada database\footnote[2]{http:$//$spaceweather.gc.ca$/$data-donnee$/$sol\_flux$/$sx-eng.php}. During the period of Antarctic measurements, the solar microwave flux varied from $65-70$ sfu's, while during our measurement period the solar flux varied from $91-93$ sfu's. If we were to correct our sky brightness measurements to match the solar activity in 1995, we would only see a reduction of 0.1 mag. This cannot be the sole explanation. Another possibility mentioned in Section \ref{sec:calibration} is the accuracy of the solid angle measurement for the photometer. It is possible that we underestimate the solid angle of the photometer, which will result in an overestimate of the sky brightness. Careful characterization in the lab can resolve this issue. It is also worth noting that the Antarctic measurements were subject to a $30-40$\% uncertainty in the J and $H$-bands, which could lead to an underestimate of the sky brightness by up to $0.3-0.4$ mags\cite{phillips99}. Although this could bring the South Pole measurements into closer agreement with ours, we still have not seen sky as dark as from the Antarctic glacial plateau.

\section{Issues Encountered and Lessons Learnt}
\label{sec:issues}
A number of issues were identified in the design of the photometer, which were mainly due to the non-ideality of the components. To make accurate radiometric measurements, the dark value of the photometer needs to be known to a high degree of accuracy. It was discovered that this dark value changed significantly with temperature. One puzzling aspect was that the photodiode should not produce any dark current because it was run in zero bias mode. Whatever bias was in the system was inherent to the readout circuit, which was designed to have a well-defined fat zero value. For sufficiently long integration times of 1s, the bias value would become negative and be clipped because it had fallen below the low end of the ADC's range. We discovered that this was because there was a current flowing through the photodiode that was in the opposite direction of the photocurrent. This current was also very sensitive to the diode temperature. It is worth noting that photodiodes do not have infinite resistances and they have a shunt resistance. The shunt resistance is inversely proportional to the diode active area and material properties. In our case, we were using a fairly large InGaAs diode, which meant its shunt resistance was only 20 M$\Omega$ at 20C. This value increases significantly to 200 M$\Omega$ when the diode temperature is $-10$C. There is an exponential relation between shunt resistance and temperature. Of course, a current would not flow through a resistor unless there was potential difference. It turns out the diode was not exactly zero-biased due to a small input offset voltage introduced by the DDC 112. This voltage is at the level of a few 10s of $\mu$V. This is enough to generate a leakage current of $\sim0.5$ pA at room temperature. However, this diminishes significantly at temperatures lower than $-10$C where it has a negligible effect. 
\par
To mitigate this issue, we used a TEC-cooled diode in order to maintain its temperature at $<-10$C at all times. Unfortunately, when we turned on the TEC we observed a rather large dark current. We managed to isolate this problem to two separate issues: (1) Leakage of the TEC driver current into the photocurrent sensing circuit; and (2) Thermal light sensitivity of the photodiode. The TEC requires a current of a few hundred mAs to adequately cool the diode. We attempted to isolate the TEC from the photo sensing circuit by driving it using a battery supply. The problem remained. This meant that there were stray impedances, most likely between the terminal posts of the photodiode, that were coupling a small portion of the TEC current into the photo sensing circuit. The second issue was originally an oversight. A typical InGaAs diode's wavelength sensitivity cuts off at 1.7$\mu$m. Since the $J$-band filter blocks all light long ward of $\sim1.3$ $\mu$m, the filter is close to 100\% emissive at wavelengths outside of its bandpass. The diode then sees blackbody emission across $2\pi$ sr at room temperature outside of the filter bandpass, and the light long ward of $1.3$ $\mu$m makes a significant contribution to the signal. Our calculations show that at an ambient temperature of 25C, we expect to see approximately 1 pA from thermal light. This was very close to the measured value in our tests. At $-10$ and $-40$C, one would expect a photocurrent of 0.02 pA and 0.001 pA from thermal photons, respectively. Therefore, the leakage current and thermal photons make negligible contributions to a photometer operating at Arctic temperatures. However, for a general purpose photometer that operates under all environmental conditions, these effects need to be carefully characterized or mitigated. 
\par
The last issue was an operational one, which was the result of ice crystals and snow collecting on the entrance window of the photometer. Since the photometer is not heated and was designed to operate at ambient temperatures, the accumulation remained on the entrance window. Fortunately, during this observing campaign, we were able to periodically check up on the photometers and physically remove any accumulation. The ultimate objective is to leave the photometer unattended over an entire winter campaign, which requires a way to automatically remove accumulation or prevent it from forming.

\section{Future Plans}
\label{sec:futureplans}
Our goal is to obtain a full winter of measurements. This would provide a sufficient baseline for solid statistics of the NIR sky brightness near PEARL. An interesting comparison might be to the Summit Station at the apex of the Greenland ice sheet. It may be that the darkest sky brightness will occur during the middle of winter when there is little to no sunlight incident on the upper atmosphere for long periods. In addition to carrying out these measurements in the Arctic, we are planning to install another version of the photometer at a mid-latitude astronomical site to obtain a long baseline comparison dataset. However, to accomplish this task, we need to resolve the thermal issues that we discussed earlier. As it stands, more work needs to be done for the photometer to function in all environmental conditions.
\par
One other area we need to improve upon is the calibration of the photometer. We have not yet accurately measured the angular response function or carried out a flux calibration across the full bandpass of the photometer. There are two possible avenues to improve the calibration. We can use an integrating sphere to generate a photometric standard and measure the response of the photometer over all angles and illuminate the entire entrance aperture. The second method requires the simultaneous measurement of the $J$-band sky background with both the photometer and a NIR camera with a $J$-band filter. The NIR camera can be first used to observe at a stellar photometric standard to carry out a flux calibration of the camera system. By observing blank sky and then a known flux calibrator, one can carry out an absolute measurement of the $J$-band sky background with the camera system. 

\section{Conclusions}

We set out to test the hypothesis that the NIR sky brightness is lower in polar regions due to suppressed OH airglow emission. Measurements in the high Antarctic plateau revealed that the $J$-band sky brightness was $2-3$ times (1 mag) fainter than the best mid-latitude sites like Mauna Kea. For this reason, we designed a rugged photometer that was suitable for operation in the High Arctic to measure the $J$-band sky background. We installed this device at PEARL near Eureka, Nunavut, Canada and carried out a short sky brightness measurement campaign in February 2012. Our measurements indicate that the $J$-band sky brightness varies between $15.5-15.9$ mag arcsec$^2,$ with a measurement uncertainty of 0.15 mag, which is dominated by systematic errors in our radiometric calibration. On our best night, we measured a fairly consistent sky brightness of $15.8\pm0.15$ mag arcsec$^2.$ This is not corrected for atmospheric extinction, which is usually $<0.1$ mag in the $J$-band at a good site. This sky brightness is comparable to the best mid-latitude sites, but is not as dark as the Antarctic measurements. We tested the idea that the increased solar activity at present times may explain the increased sky brightness. However, we discovered that using known correlations between OH line flux and 10.7cm solar flux to correct for this effect can only account for a 0.1 mag decrease in the sky brightness. One other possible explanation is that we simply do not have sufficient statistics on the variation of the sky brightness in the Arctic due to the short span of our measurements. Finally, we may underestimate the solid angle of the photometer, which means our measured sky brightness value is actually somewhat higher than the true value. Future improvements in our calibration techniques and a winter-long sky brightness measurement campaign should resolve these issues. 

\label{sec:conclusions}

\acknowledgments     
 
S.S. was supported by the Dunlap Institute through the Dunlap fellowship program. S.S. would like to thank Bill Reeve and Mark Anderson of Lockheed Martin Corporation for our long discussions about making precise pico amp level current measurements and the intricacies of op amps. S.S. would also like to thank Adam Palmentieri of Hamamatsu Corporation for his assistance in troubleshooting spurious dark current in the readout circuit. We are grateful to the Canadian Network for the Detection of Atmospheric Change, particularly Pierre Fogal and James Drummond for their assistance in deploying the photometer on the PEARL roof. The Eureka weather station is operated by Environment Canada, and E.S. and W.N. greatly appreciate the warm hospitality of their staff during our \"observing run." This research uses data from Space Weather Canada solar flux database.

\bibliography{report}   
\bibliographystyle{spiebib}   

\end{document}